\documentclass[journal]{IEEEtran}
\usepackage{cite}
\usepackage{textcomp}
\usepackage{url}
\usepackage{verbatim}
\usepackage{amsmath, amsfonts, amssymb}
\usepackage{bm}
\usepackage{graphicx}
\usepackage{subcaption}
\usepackage{tikz}
\usepackage{multirow}
\usetikzlibrary{arrows.meta, positioning}
\usepackage{tikz-cd}
\usepackage{tabularx}
\usepackage{array}
\usepackage{booktabs}
\usepackage{makecell}

\usepackage{enumitem}
\usepackage{algorithm}
\usepackage{algpseudocode}
\usepackage{stfloats}
\usepackage{xcolor} % added for highlighting changes
\usepackage{hyperref}
\hypersetup{
    colorlinks=true,
    linkcolor=blue,
    filecolor=magenta,
    urlcolor=cyan,
    citecolor=green,
    pdftitle={Your Title},
    pdfauthor={Your Name},
}

\def\BibTeX{{\rm B\kern-.05em{\sc i\kern-.025em b}\kern-.08em
    T\kern-.1667em\lower.7ex\hbox{E}\kern-.125emX}}

\newcounter{remark} % Global counter, not reset per section
\newcommand{\remark}[1]{%
  \stepcounter{remark}%
  \noindent\textbf{Remark \arabic{remark}:} #1\par
}

\begin{document}

\title{Flow-based Polynomial Chaos Expansion for Uncertainty Quantification in Power System Dynamic Simulation}

\author{
\IEEEauthorblockN{Le Fang, Wangkun Xu, Fei Teng\\}
}

\maketitle

\begin{abstract}

The large-scale integration of renewable energy sources introduces significant operational uncertainty into power systems. Although Polynomial Chaos Expansion (PCE) provides an efficient tool for uncertainty quantification (UQ) in power system dynamics, its accuracy depends critically on the faithful representation of input uncertainty, an assumption that is oftern violated in practice due to correlated, non-Gaussian, and otherwise complex data distributions. In contrast to purely data-driven surrogates that often overlook rigorous input distribution modelling, this paper introduces flow-based PCE, a unified framework that couples expressive input modelling with efficient uncertainty propagation. Specifically, normalising flows are employed to learn an invertible transport map from a simple base distribution to the empirical joint distribution of uncertain inputs, and this map is then integrated directly into the PCE construction. In addition, the Map Smoothness Index (MSI) is introduced as a new metric to quantify the quality of the learned map, and smoother transformations are shown to yield more accurate PCE surrogates. The proposed Flow-based PCE framework is validated on benchmark dynamic models, including the IEEE 14-bus system and the Great Britain transmission system, under a range of uncertainty scenarios.
\end{abstract}


\section{Introduction}

\subsection{Background and Motivation}

The global push towards decarbonization fundamentally redefines modern power systems through the large-scale integration of renewable energy sources (RES) \cite{bird_integrating_2013}. While essential for sustainability, RES introduces significant operational challenges because their output is inherently variable and only partially predictable \cite{foley_current_2012}. In parallel, uncertainty is rising on the demand side, driven by new technologies such as electric vehicles and smart loads \cite{siano_demand_2014}. As a result, modern power systems operate under more stressed conditions with a much higher degree of uncertainty than in the past.

Such uncertainties on both the generation and demand sides propagate through the nonlinear power system dynamic models, affecting the outcomes of critical operational tasks such as transient stability assessment (TSA), security monitoring, and the preventive control design. Traditional deterministic analysis, which often relies on a limited set of robust worst-case scenarios, is no longer sufficient to capture the full spectrum of possible system behaviours \cite{Milanovic_ProbStab2017}. Incomplete or biased assessment of dynamic behaviour can lead to incorrect judgements on system stability, potentially causing system-wide blackout. These realities motivate a shift from deterministic analyses toward probabilistic dynamic assessment, and hence the development of computationally efficient uncertainty quantification (UQ) tools \cite{wu_probabilistic_1983}.
% \cite{Billinton_ProbabilisticAssessment,Billinton_ProbabilisticIndex,Kuruganty_ProtectionSystem,wu_probabilistic_1983}

\subsection{Uncertainty Quantification}

The field of Uncertainty Quantification (UQ), originally developed in fields like reliability and civil engineering, provides a formal framework for the analysis of systems under uncertainty. A complete UQ study is typically broken down into three main tasks: characterization of the computational model, quantification of the source uncertainty, and propagation of this uncertainty through the model \cite{sudret_uncertainty_2007}.

\subsubsection{Computational Model}

The computational model can be viewed as an abstract function, $\mathcal{M}$, that maps input parameters, $\bm{\xi}$, to a response, $\mathbf{Y} = \mathcal{M} (\bm{\xi})$. In many real scenarios, it is treated as a black box model.

\subsubsection{Source Uncertainty Quantification}

Characterizing the uncertainty in a system's input parameters is the foundational step. This uncertainty can be broadly categorized as \textit{aleatoric}, arising from inherent randomness in a phenomenon, and \textit{epistemic}, resulting from a lack of knowledge that can potentially be reduced with more data. In recent years, the machine learning community has developed powerful tools for this task under the umbrella of deep generative modelling. State-of-the-art approaches \cite{bond-taylor_deep_2022}, including variational autoencoders, autoregressive models, diffusion models, and normalising flows, have demonstrated the ability to learn highly complex, multimodal, and correlated distributions directly from data, moving far beyond the limitations of classical parametric models.

\subsubsection{Uncertainty Propagation}

Once input uncertainty has been modelled, the next challenge is to propagate it through the computational model and quantify output uncertainty. While Monte Carlo simulation is the most straightforward approach, its computational cost is often prohibitive for complex power-system models. Spectral methods, particularly Polynomial Chaos Expansion (PCE), have therefore emerged as efficient alternatives. PCE approximates the model response as a polynomial expansion in random inputs, enabling analytical estimates of statistical moments and sensitivities with far fewer model evaluations than Monte Carlo \cite{soize_physical_2004,crestaux_polynomial_2009,ernst_convergence_2012}. Recent research in PCE has focused on improving performance in high-dimensional settings \cite{lataniotis_data-driven_2019}, handling dependent inputs \cite{feinberg_multivariate_2018,jakeman_polynomial_2019,rahman_polynomial_2018}, and developing adaptive, data-driven variants \cite{torre_data-driven_2019}.

Beyond PCE, moment-based approximations (e.g., cumulant) are computationally efficient but may fail for strongly non-linear or multimodal responses \cite{sudret_uncertainty_2007}. In addition, Gaussian Process (GP) emulators provide Bayesian surrogates with predictive uncertainty, but they are fundamentally designed to quantify \emph{model} (epistemic) uncertainty about the surrogate itself, rather than to propagate \emph{input} (aleatoric) uncertainty through a deterministic simulator. While the GP posterior mean can serve as a deterministic surrogate, it discards the core advantage of GPs and still requires Monte-Carlo sampling to obtain output statistics, which reintroduces cost and scaling challenges in higher dimensions~\cite{tan_gaussian_2026}. In contrast, PCE directly encodes input-to-output uncertainty in its coefficients, making it a more natural choice for power system UQ with input uncertainties such as renewable profile and forecast error.

As a result, PCE techniques have been increasingly applied in power system analysis, including probabilistic power flow \cite{Ni_powerflow}, dynamic simulation \cite{xu_propagating_2019,qiu_nonintrusive_2020,ye_uncertainty_2021} and optimal power flow \cite{metivier_efficient_2019} under uncertain renewables. However, a fundamental gap remains between uncertainty modelling and uncertainty propagation. In particular, conventional PCE requires \emph{independent} input random variables to construct orthogonal polynomial bases. This requirement is often violated in practice, as empirical evidence shows that renewable uncertainties or forecast errors exhibit strong spatial and temporal dependence, as well as non-Gaussian heavy-tailed behaviour \cite{fang2018spatial,schafer2018nonGaussian,bruninx2014statistical}. Although parametric approaches such as Gaussian copulas \cite{fan_uncertainty_2021} can be used to model dependent uncertainty sources, they often struggle to capture high-dimensional or multimodal distributions. Therefore, there is a clear need for a unified framework that can i) model complex uncertainties sources, and ii) transform these uncertainties into independent latent embeddings that are compatible with the polynomial basis construction in conventional PCE.

% that integrates data-driven source characterisation with classical uncertainty propagation methods.
% In practice, this mismatch is often bypassed by assuming independence, sacrificing fidelity for tractability. When dependence is incorporated, it is typically modelled using parametric tools such as Gaussian copulas \cite{fan_uncertainty_2021}, which can struggle to capture complex or multimodal dependence structures. Consequently, there remains a need for a unified framework that integrates data-driven source characterisation with classical uncertainty propagation methods.

% Data-driven extensions of PCE do not resolve this gap. Approaches that learn nonparametric marginals via KDE \cite{torre_data_driven_pce_2019} or construct bases from empirical moments (arbitrary PCE) \cite{oladyshkin_apc_2012} still assume independent inputs or defer to copulas for dependence \cite{wang_apc_plf_2019}, leaving G1 unaddressed. Furthermore, these methods introduce a second gap (\textbf{G2}): compared with generalised PCE, they offer limited convergence-rate guidance, lack a practical theoretical link between the quality of the input model and end-to-end surrogate accuracy, and can suffer from computational instability when moment estimates are noisy or the basis construction is ill-conditioned.

\subsection{Contributions}

The main contributions of this work are summarised as follows:

\begin{itemize}[leftmargin=*]
    \item \textbf{Joint modelling and propagation framework.} We propose a unified Flow‐based PCE framework that learns the \emph{full joint distribution} of uncertain inputs using a normalising flow and, through its invertible transport map, transforms correlated physical variables into independent latent variables. This eliminates the need for prescribed parametric distributions on uncertainties, and enables a \emph{single} workflow that couples input modelling with efficient uncertainty propagation.

    \item \textbf{Theoretical analysis and practical guidance.}
    By mapping arbitrary dependent inputs to independent Gaussian latent variables, the normalising flow enables the use of classical gPC with Hermite polynomial bases in the latent space, thereby inheriting its well-established convergence properties. To make this connection actionable, we derive a guidance-oriented generalisation bound that relates the end-to-end surrogate error to (i) the distribution mismatch quantified by the Wasserstein distance; (ii) the smoothness of the learned transport map; and (iii) the sample size. Based on this analysis, we provide practical recommendations for selecting the flow architecture and the PCE degree.
    
    \item \textbf{Scalability on large power network.} 
    Power system dynamic simulation results show that the proposed flow-based PCE consistently outperforms state-of-the-art PCE methods based on copula-based input modelling under both \emph{unimodal} and \emph{multimodal} input distributions. A large-scale Great Britain transmission system case further demonstrates the scalability of the proposed method, which maintains high accuracy under high-dimensional inputs, whereas the copula-based approach degrades substantially.
\end{itemize}

This paper is organized as follows. Section \ref{sec:prob_formulate} provides basic UQ formulation in terms of power system dynamic simulation. Section \ref{sec:pce_background} provides the background of the PCE method. Section \ref{sec:flow_pce} presents the proposed flow-based PCE. Case studies are described in Section \ref{sec:case_study} while this paper concludes in Section \ref{sec:conclusion}.

\section{Problem Formulation}\label{sec:prob_formulate}

The overall problem of propagating uncertainty through a power system dynamics model can be broken down into two sub-problems.

First, the sources of uncertainty must be characterized. It is assumed the uncertainty originates from a set of $M$ physical parameters, represented by a random vector $\bm{\xi} \in \mathbb{R}^M$, whose joint probability density function (PDF), $f_{\bm{\Xi}}(\bm{\xi})$, is unknown. The goal of this step is to construct an accurate representation of $f_{\bm{\Xi}}$, typically by applying statistical inference or density estimation techniques to a given set of observations or data, $\{\bm{\xi}^{(1)}, \dots, \bm{\xi}^{(N)}\}$. It is worth noting that in prior studies in power systems, this step is often bypassed by presuming a known and simplified distribution. In reality, however, such distributions are only conceptual constructs and should be inferred from observed data.

Once the input uncertainty is characterized, the second step is to propagate it through the system model. The system response is governed by a computational model, which is denoted abstractly as a function $\mathcal{M}$. This model takes the random input vector $\bm{\xi}$ and produces a random output $\mathbf{Y}$:

\begin{equation}\label{eq:dynamic_model}
    \mathbf{Y} = \mathcal{M}(\bm{\xi}), \quad \bm{\xi}\sim f_{\bm{\Xi}} (\bm{\xi})
\end{equation}
Depending on the property of $\mathcal{M}$, $\mathbf{Y}$ can be a random scalar or vector. 

Directly determining the full probability distribution of the output $\mathbf{Y}$ is intractable, as it requires propagating the input density $f_{\bm{\Xi}}(\bm{\xi})$ through the complex, non-linear model $\mathcal{M}$. The most direct numerical approach to this problem is Monte Carlo (MC) simulation, which involves repeatedly evaluating the model for a large number of random input samples. However, due to the high computational expense of the underlying model, the direct use of MC is also infeasible.

Considering power system dynamic simulations, the governing model $\mathcal{M}$ is a set of Differential-Algebraic Equations (DAEs), which, together with their initial state, are given by:
\begin{equation}
\begin{cases}
    \dot{\mathbf{x}} = f(\mathbf{x}, \mathbf{y}, \bm{\eta}) \\
    \mathbf{0} = g(\mathbf{x}, \mathbf{y}, \bm{\eta}) \\
    (\mathbf{x}(0), \mathbf{y}(0), \bm{\eta}(0)) = (\mathbf{x}_0, \mathbf{y}_0, \bm{\eta}_0)
\end{cases}
\label{eq:dae_system_full}
\end{equation}
where $\mathbf{x}$ represents the state (dynamic) variables, $\mathbf{y}$ the algebraic variables, and $\bm{\eta}$ a vector of uncertain system parameters. While often implicitly embedded in $f$ and $g$, $\bm{\eta}$ is explicitly taken to represent physical quantities such as control settings and voltage references. In this study, we consider the primary source of uncertainties to be the power injections from RES, thus setting our random input vector $\bm{\xi} \equiv \bm{\eta}$. The dynamic simulation workflow dictates how this uncertainty propagates.
First, a specific realization of the random power injections $\bm{\xi}$ is used as an input to a power flow calculation. The results of the power flow (i.e., bus voltages and angles) are then used to solve an initialization problem, which determines the consistent initial conditions $(\mathbf{x}_0(\bm{\xi}), \mathbf{y}_0(\bm{\xi}), \bm{\xi})$ for the DAE system. Consequently, the whole UQ process for power system dynamics \eqref{eq:dae_system_full} is denoted as $\mathcal{M}$, and $\mathbf{Y} = \mathcal{M}(\bm{\xi})$. Note that the output value $\mathbf{Y}$ can be any quantity from the simulation results, such as the trajectory of a specific state variable $x_i(t, \bm{\xi})$ or an algebraic variable $y_j(t, \bm{\xi})$.
% , or any functional thereof.

% \remark{The formulation above describes a probabilistic Initial Value Problem. Other ways to introduce stochasticity, e.g. modelling disturbances with continuous-time random processes, lead to Stochastic Differential-Algebraic Equations and require dimensionality reduction before applying PCE. }

\section{Theoretical Background on PCE}\label{sec:pce_background}

The main target of PCE is to learn a parametric approximation of $\mathcal{M}(\bm{\xi})$ in \eqref{eq:dynamic_model} which allows statistical indices of the model output to be obtained directly from the learnt parameters. Compared to MC simulation, PCE requires fewer samples to obtain accurate and numerically stable results. This section introduces PCE for independent uncertainty propagation.

\subsection{Polynomial Chaos Expansion for Independent Input}

Consider a model $ \mathcal{M}: \mathbb{R}^M \to \mathbb{R}$ with output $Y = \mathcal{M}(\bm{\xi})$, where the input is a random vector $\bm{\xi} = (\xi_1, \dots, \xi_M)$ with joint PDF, $f_{\bm{\Xi}}(\bm{\xi})$. The model must have a finite second moment, $\mathbb{E}[\mathcal{M}(\bm{\xi})^2] < \infty$, which allows the model function $\mathcal{M}$ to be treated as an element of the Hilbert space $\mathcal{H} = L^2_{P_{\bm{\Xi}}}(\mathbb{R}^M, \mathbb{R})$. This space of square-integrable functions is endowed with the inner product:

\begin{equation}\label{eq:inner_prod}
    \langle u, v \rangle_{\mathcal{H}} = \int_{\mathbb{R}^M} u(\bm{\xi}) v(\bm{\xi}) f_{\bm{\Xi}}(\bm{\xi}) \,d\bm{\xi}
\end{equation}

The standard PCE construction is particularly convenient when the input variables $\{\xi_i\}$ are \emph{mutually independent}. The joint PDF then factors into the product of marginals,

\begin{equation}\label{eq:joint_pdf}
f_{\bm{\Xi}}(\bm{\xi}) = \prod_{i=1}^{M} f_{\Xi_i}(\xi_i)
\end{equation}
Under \eqref{eq:joint_pdf}, the space $\mathcal{H}$ becomes a tensor product of one-dimensional spaces, $\mathcal{H} = \bigotimes_{i=1}^{M} \mathcal{H}_i$. Each space $\mathcal{H}_i$ can be constructed by an orthonormal basis of polynomials $\{\psi_k^i\}_{k \in \mathbb{N}}$ with respect to the weight function $f_{\Xi_i}(\xi_i)$. A multivariate orthonormal basis for $\mathcal{H}$ is then obtained via tensor products of the univariate basis. Specifically, for a multi-index $\bm{\alpha} = (\alpha_1, \dots, \alpha_M) \in \mathbb{N}^M$, define the basis function

\begin{equation}\label{eq:basis_construction}
    \Psi_{\bm{\alpha}}(\bm{\xi}) = \prod_{i=1}^{M} \psi_{\alpha_i}^i(\xi_i)
\end{equation}
The collection $\{\Psi_{\bm{\alpha}}: \bm{\alpha}\in\mathbb{N}^M\}$ forms a complete orthonormal basis of $\mathcal{H}$.

Accordingly, any function $h(\bm{\xi}) \in \mathcal{H}$ can be uniquely represented by its series expansion onto this basis, known as the \textit{Polynomial Chaos Expansion} (PCE):

\begin{equation}
    h(\bm{\xi}) = \sum_{\bm{\alpha} \in \mathbb{N}^M} h_{\bm{\alpha}} \Psi_{\bm{\alpha}}(\bm{\xi})
\end{equation}
and the spectral coefficients $h_{\bm{\alpha}}$ are found via orthogonal projection defined by the inner product \eqref{eq:inner_prod}:

\begin{equation}
    h_{\bm{\alpha}} = \langle h, \Psi_{\bm{\alpha}} \rangle_{\mathcal{H}} = \int_{\mathbb{R}^M} h(\bm{\xi}) \Psi_{\bm{\alpha}}(\bm{\xi}) f_{\bm{\Xi}}(\bm{\xi}) \,d\bm{\xi}
    \label{eq:projection}
\end{equation}

In literature, this approach is known as the generalized Polynomial Chaos (gPC) expansion \cite{xiu_numerical_2010} when the marginal distributions $f_{\Xi_i}$ belong to the Askey–Wiener family of distributions. In such cases, the corresponding orthogonal polynomials are well-known classical polynomials (e.g., Hermite for Gaussian inputs and Legendre for uniform inputs), greatly simplifying practical application. For brevity, we omit the standard correspondence table and refer readers to \cite{xiu_numerical_2010}.

\remark{The scalar notation for the output can be easily modified for the vector cases where a series of Hilbert space for each discrete time steps can be defined.}

\subsection{Coefficient Estimation}

Since modifying the computational model $\mathcal{M}$ is often infeasible, our focus is on \emph{non-intrusive} methods for estimating the PCE coefficients. These methods treat the model as a black box. Two primary categories of non-intrusive techniques exist: projection and regression. Projection methods approximate the true integral in \eqref{eq:projection} using techniques like Monte Carlo simulation or numerical quadrature. In this work, we adopt the regression-based approach, as it provides a flexible framework where regularization can be naturally incorporated to prevent overfitting and improve the convergence rate of the PCE surrogate.

For practical computation, considering the infinite polynomials in $\mathbb{N}^M$ is intractable. Therefore, a truncated basis set is considered by selecting all multivariate polynomials $\Psi_{\bm{\alpha}}$ whose total degree is less than or equal to a chosen integer $p$:

\begin{equation}
    \mathcal{A} = \left\{ \bm{\alpha} \in \mathbb{N}^M : |\bm{\alpha}| = \sum_{i=1}^{M} \alpha_i \le p \right\}
    \label{eq:trunc_set}
\end{equation}
The number of terms in this truncated basis is given by $P = \text{card}(\mathcal{A}) = \binom{M+p}{p}$. 
The model response can then be expressed as the sum of a truncated PCE and a truncation error $\epsilon_p$:

\begin{equation}
    Y = \mathcal{M}(\bm{\xi}) = \sum_{\bm{\alpha} \in \mathcal{A}} h_{\bm{\alpha}} \Psi_{\bm{\alpha}}(\bm{\xi}) + \epsilon_p
    \label{eq:pce_truncated_sum}
\end{equation}
To simplify the notation, the coefficients and the polynomials are grouped into vector form $\mathbf{h} = (h_{\bm{\alpha}})_{\bm{\alpha} \in \mathcal{A}} \in \mathbb{R}^P$ and design matrix $\bm{\Psi}(\bm{\xi}) = (\Psi_{\bm{\alpha}}(\bm{\xi}))_{\bm{\alpha} \in \mathcal{A}} \in \mathbb{R}^P$, respectively.  

A natural way to compute the unknown coefficient vector $\mathbf{h}$ is to formulate the problem as a linear regression. The optimal set of coefficients, denoted by $\hat{\mathbf{h}}$, is the one that minimizes the mean squared error of the approximation. This leads to the following least-squares minimization problem:

\begin{equation}
    \hat{\mathbf{h}} = \arg\min_{\mathbf{h} \in \mathbb{R}^P} \mathbb{E}\left[ \left(\mathcal{M}(\bm{\xi}) - \mathbf{h}^\top \bm{\Psi}(\bm{\xi})\right)^2 \right] + \lambda\,\mathcal{R}(\mathbf{h})
    \label{eq:least_squares_problem}
\end{equation}
In practice, the expectation in \eqref{eq:least_squares_problem} is replaced by an empirical average over $N_s$ samples, i.e., $\{\bm{\xi}^{(n)}\}_{n=1}^{N_s}$ (typically drawn from $f_{\bm{\Xi}}$ using random or quasi-random designs), leading to a discrete least-squares problem,
\begin{equation}\label{eq:least_squares_problem_1}
\hat{\mathbf{h}}=\arg\min_{\mathbf{h}\in\mathbb{R}^P}\ \frac{1}{N_s}\sum_{n=1}^{N_s}\left(y^{(n)}-\mathbf{h}^\top\bm{\Psi}(\bm{\xi}^{(n)})\right)^2 + \lambda\,\mathcal{R}(\mathbf{h})
\end{equation}
where $y^{(n)}=\mathcal{M}(\bm{\xi}^{(n)})$. For numerical stability, the regression is typically taken overdetermined with $N_s$ larger than $P$ (e.g., $N_s \geq 2P$ as a rule of thumb), and $\mathcal{R}(\mathbf{h})$ may be an $\ell_2$ penalty (ridge) for conditioning or an $\ell_1$ penalty (LASSO) for sparsity.
While orthonormal polynomials imply favourable conditioning \emph{in expectation} under ideal sampling, finite-sample stability ultimately depends on the sampling design and the conditioning of the design matrix.

% This approach is often referred to as a \textit{probabilistic collocation method}. The primary design choice for this method is the selection of the collocation points. While there are various strategies, these points are traditionally drawn from random or quasi-random sequences (e.g., Sobol) or chosen as the nodes of a numerical quadrature rule. For a stable solution, the regression is typically overdetermined by choosing a number of samples $N_s$ that is significantly larger than the number of unknown coefficients $P$ (e.g., $N_s \ge 2P$). Thanks to the orthogonal polynomial basis, the coefficients are uniquely identifiable and numerically stable to compute.

\subsection{Post Processing of PCE Coefficients}
A primary advantage of the PCE method is that once the coefficients $\mathbf{h}$ are computed, much statistical information about the model output can be retrieved efficiently. This post-processing is performed directly on the PCE surrogate model, $\mathcal{M}_{\text{PCE}}(\bm{\xi}) = \mathbf{h}^\top \bm{\Psi}(\bm{\xi})$, and can be done either analytically or with minimal computational effort. For example, due to the orthonormality of the polynomial basis functions $\Psi_{\bm{\alpha}}(\bm{\xi})$, the mean and variance of the output variable $Y$ are available in closed form directly from the coefficients. Assuming the basis is constructed such that $\Psi_{\bm{0}} = 1$ and $\mathbb{E}[\Psi_{\bm{\alpha}}] = 0$ for all $\bm{\alpha} \neq \bm{0}$, the mean and variance are given by:

\begin{subequations}\label{eq:pce_statistic}
    \begin{equation}
        \mu_Y^{\text{PCE}} \equiv \mathbb{E}[Y] = h_{\bm{0}} \label{eq:pce_mean}
    \end{equation}
    \begin{equation}\label{eq:pce_variance}
        (\sigma_Y^{\text{PCE}})^2 \equiv \text{Var}[Y] = \sum_{\bm{\alpha} \in \mathcal{A} \setminus \{\bm{0}\}} h_{\bm{\alpha}}^2
    \end{equation}
\end{subequations}
where $h_{\bm{0}}$ is the coefficient corresponding to the zero-degree polynomial basis function and the sum for the variance is over all non-constant basis functions in the truncated set $\mathcal{A}$ in \eqref{eq:trunc_set}.

Beyond these basic moments, the PCE surrogate is invaluable for more advanced reliability and sensitivity studies. The corresponding metrics can be obtained by applying their definitions directly to the PCE approximation. Some quantities, such as global sensitivity metrics like Sobol' indices, can be calculated analytically from the PCE coefficients as well \cite{sudret2008global}.

% For other analyses, such as estimating failure probabilities, one can use Monte Carlo methods. The key advantage is that the sampling is performed on the analytical and computationally cheap PCE surrogate, not the original model. The cost of such a simulation is usually negligible compared to the computational effort required to build the PCE model in the first place.

\section{Normalising Flow-based PCE}\label{sec:flow_pce}

\subsection{Dependent Inputs and Mapping}
\subsubsection{Handling Input Dependence} 

The tensor–product construction of multivariate polynomial bases only yields an orthogonal PCE when the random inputs are independent. For dependent inputs, directly using product polynomials destroys orthogonality and can produce erroneous statistics. A naive workaround is to ignore dependence and build a PCE from the marginals, but this fails to capture covariance and yields unreliable variance and reliability estimates. To address this, several strategies exist \cite{feinberg_multivariate_2018}: (i) mapping methods that transform dependent variables to independent latent variables before constructing a standard PCE; (ii) dominating–measure approaches that build a PCE with respect to a simpler product measure and correct for discrepancies; and (iii) explicit construction of multivariate orthonormal polynomials via Gram–Schmidt procedures \cite{jakeman_polynomial_2019}. The latter two options often become computationally prohibitive or numerically unstable in high dimensions. Consequently, we adopt the mapping strategy.

% both methods are known to be numerically unstable 

\subsubsection{Mapping Method}
The goal is to find an invertible transformation $T$ such that
$\bm{\xi}=T(\mathbf{Z})$. The system response can then be approximated as
$\mathcal{M}(\bm{\xi})\approx\sum_{\bm{\alpha}} h_{\bm{\alpha}}\Psi_{\bm{\alpha}}(\mathbf{Z})$
using a standard PCE in the latent space. The invertibility also ensures the sampling from latent space
for evaluation purposes.

In practice, one common way to obtain such a map $T$ is to first construct an invertible transformation in
the reverse direction, i.e., a map that sends the physical variables to independent latents,
$\mathbf{Z}=S(\bm{\xi})$. If $S$ is invertible, then $T=S^{-1}$ provides the desired latent-to-data map.

Classical choices for constructing $S$ include the Rosenblatt transform \cite{rosenblatt1952remarks}, which
uses a sequence of conditional CDFs, and the Gaussian-copula-induced Nataf transform \cite{noh_reliability-based_2009},
which assumes the dependence structure is well represented by a Gaussian copula.
By Sklar’s theorem, a joint CDF can be equivalently represented as
\begin{equation}\label{eq:nataf_1}
F_{\boldsymbol{\Xi}}(\boldsymbol{\xi}) \;=\; C\!\big(F_{\Xi_1}(\xi_1),\ldots,F_{\Xi_M}(\xi_M)\big),
\end{equation}
In practice, the marginal CDF $F_{\Xi_i}$ is first fitted from data and then a copula $C$ is chosen to encode dependence. When the copula is taken to be \emph{Gaussian}, the resulting construction induces the Nataf transformation \cite{noh_reliability-based_2009}, an isoprobabilistic map that yields independent Gaussian latents and is especially convenient for PCE. The Gaussian copula with correlation matrix $\mathbf{R}_0$ is defined as
\begin{equation}
C_{\mathbf{R}_0}(\mathbf{u}) \;=\; \Phi_{\mathbf{R}_0}\!\big(\Phi^{-1}(u_1),\ldots,\Phi^{-1}(u_M)\big),
\end{equation}
where $\Phi_{\mathbf{R}_0}$ is the CDF of a centred multivariate normal with correlation \(\mathbf{R}_0\) and \(\Phi^{-1}\) is the standard univariate normal quantile.

% The Nataf transformation \(T_{\mathrm{nat}}=T_2\circ T_1\) is defined by first mapping the marginal CDF into the latent vector $\mathbf{Z}'$ and then decorrelates into independent Gaussian variables $\mathbf{Z}$,
The Nataf map \(S_{\mathrm{nat}} = S_2\circ S_1\) consists of two steps:
\begin{subequations}
\label{eq:nataf}
\begin{align}
    S_1: \bm{\xi} \mapsto \mathbf{Z}' &= 
    \begin{pmatrix}
        \Phi^{-1}(F_{\Xi_1}(\xi_1)) \\
        \vdots \\
        \Phi^{-1}(F_{\Xi_M}(\xi_M))
    \end{pmatrix} \\
    S_2: \mathbf{Z}' \mapsto \mathbf{Z} &= \mathbf{L}_0^{-1} \mathbf{Z}'
\end{align}
\end{subequations}
where $\mathbf{R}_0=\mathbf{L}_0\mathbf{L}_0^{\top}$ is the Cholesky factorization. The first step $S_1$ enforces standard normal \emph{marginals} for $\mathbf{Z}'$; under the Gaussian-copula assumption, $\mathbf{Z}'$ is jointly Gaussian with correlation $\mathbf{R}_0$. The second step $S_2$ whitens $\mathbf{Z}'$ to produce $\mathbf{Z}\sim \mathcal{N}(\mathbf{0},\mathbf{I})$. Since both $S_1$ and $S_2$ are invertible, the desired latent-to-data map is given by
$T_{\mathrm{nat}} = S_{\mathrm{nat}}^{-1} = S_1^{-1}\circ S_2^{-1}$, so that $\bm{\xi}=T_{\mathrm{nat}}(\mathbf{Z})$.

In this paper, using copulas to model dependence (with fitted marginals) and then building the PCE on the associated standardised variables is referred to as the \emph{Copula–PCE} method.

\remark{Copula methods decouple marginals and dependence: one first fits the univariate CDFs and then selects a copula family to model dependence. While attractive in a data‐driven context, the copula family choice is a substantive modelling decision and can bias the implied joint distribution when the assumed family is misspecified.}

\subsection{Input Uncertainty Characterisation via the Flow}

% The assumptions considered in Copula-PCE can be restrictive in high-dimensional, multimodal, or strongly non-Gaussian settings.
% In this work, we replace fixed parametric maps with a learned \emph{normalising flow} (introduced later), which provides a flexible, invertible, and smooth transport from the observed input distribution to an independent latent distribution. This enables the use of the well-established PCE machinery in latent space while learning complex dependence directly from data.

As discussed in the previous section, the key challenge of the mapping-based PCE for dependent inputs is to find a map or sequence of maps from the dependent input distribution $\bm{\xi}$ with a complex joint distribution towards an independent base distribution (e.g., Gaussian, uniform, etc.). Meanwhile, classical dependence modelling tools, such as copulas, are often limited by restrictive, pre-specified parametric forms, which may fail to accurately capture complex or non-standard dependence structures present in the data. Therefore, we propose an end-to-end Flow-based PCE framework that learns an expressive transport map directly from data, thereby standardising the inputs while preserving the empirical joint distribution. Concretely, we replace fixed-structure transformations (e.g., Rosenblatt/Nataf maps) with a highly expressive invertible neural network. Among several invertible architectures \cite{buffelli2024exact}, we adopt normalising flows, which provide (i) a tractable likelihood for learning the input density and (ii) an explicit bijection that transforms samples from the observed input distribution to an independent base distribution, enabling standard tensor-product PCE in latent space.

To start, let $\mathbf{z}$ be a random vector with a known PDF $p_Z(\mathbf{z})$ and apply an invertible and differentiable transformation $T:\mathbb{R}^M\rightarrow\mathbb{R}^M$ into $\bm{\xi} = T(\mathbf{z})$. Then the PDF of $\bm{\xi}$, denoted $p_\Xi(\bm{\xi})$, is given by the change-of-variable formula,

\begin{equation}
    p_\Xi(\bm{\xi}) = p_Z(T^{-1}(\bm{\xi})) \cdot \left| \det \left( J_{T^{-1}}(\bm{\xi}) \right) \right|
    \label{eq:change_of_vars}
\end{equation}
where $J_{T^{-1}}(\bm{\xi})$ is the Jacobian matrix of the inverse transformation $T^{-1}$ evaluated at $\bm{\xi}$. The necessity of invertible and differentiable $T$ becomes evident to represent a well-defined distribution.

\begin{figure*}[!t]
  \centering
  %----------------------------------------------
  \begin{tikzpicture}[
      node distance   = 2.5cm,
      every node/.style={font=\small},
      var/.style      ={inner sep=4pt, minimum width=13mm, minimum height=16mm,
                        draw, rounded corners=2pt, align=center},
      base/.style     ={var, fill=green!15!white},
      mid/.style      ={var, fill=pink!25!white},
      target/.style   ={var, fill=blue!15!white},
      arr/.style      ={<->, thick}
    ]

    % --- nodes ---
    \node[base]   (z0)  {$\mathbf z_0$};
    \node[mid]    (z1)  [right=of z0] {$\mathbf z_1$};
    \node[mid]    (z2)  [right=of z1] {$\mathbf z_2$};
    \node[mid]    (zk)  [right=of z2] {$\mathbf z_{K-1}$};
    \node[target] (xi)  [right=of zk] {$\bm{\xi}$};

    % --- bidirectional arrows with dual labels ---
    \draw[arr] (z0)  -- node[above]{$T_1^{-1}$} node[below]{$T_1$} (z1);
    \draw[arr] (z1)  -- node[above]{$T_2^{-1}$} node[below]{$T_2$} (z2);
    \draw[arr] (z2)  -- node[above]{$\cdots$}   node[below]{$\cdots$} (zk);
    \draw[arr] (zk)  -- node[above]{$T_K^{-1}$} node[below]{$T_K$} (xi);
  \end{tikzpicture}
  %----------------------------------------------
   \caption{A normalizing flow can be visualized as a sequence of bijective transformations. A sample drawn from the base distribution $\mathbf{z}_0$ is mapped to the complex target distribution $\bm{\xi}$ through $K$ invertible functions $T_1, \dots, T_K$. Since each $T_i$ is bijective, the transformations are reversible, illustrating that the flow operates both as a generative model (from latent to data space) and as an exact mapping method (from data to latent space) without information loss.}
  \label{fig:flow_diagram}
\end{figure*}
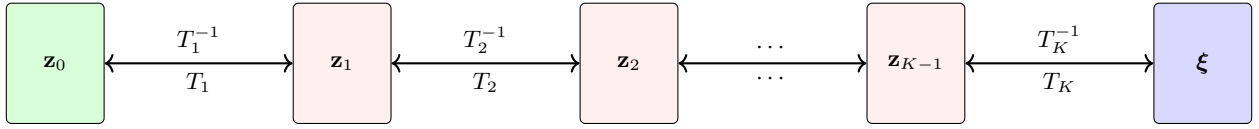

Normalising flows are a class of generative models that operationalize the change-of-variables principle for density learning. As shown by Fig.~\ref{fig:flow_diagram}, it transforms a simple base distribution $\mathbf{z}$ (e.g. standard normal) into a complex target distribution through a sequence of invertible and differentiable transformations.
% This has found wide application in density estimation, generative modeling, and variational inference \cite{papamakarios2021normalizing, kobyzev2020normalizing}.
The flow is constructed as a composition of $K$ invertible maps, $T = T_K \circ \dots \circ T_1$ which creates a bijective mapping between the base space and the target space. The term ``flow" refers to the trajectory that samples from the base distribution follow as they are gradually transformed. The term ``normalising" often refers to the inverse direction, which transforms data from a complex distribution back to a simple, standard normal base distribution. For this construction to be practical, each transformation block $T_k$ must satisfy two key conditions: (i) it must be easily invertible, and (ii) its Jacobian determinant must be easy to compute. These maps are typically parameterized by neural networks to maximize their capacity.

To train a normalising flow on a dataset $\{\bm{\xi}^{(i)}\}_{i=1}^N$, the standard approach is to maximize the log-likelihood of the data. This is equivalent to minimizing the forward Kullback-Leibler (KL) divergence between the empirical data distribution and the model distribution. The log-likelihood of a single data point $\bm{\xi}$ under the flow model is derived from \eqref{eq:change_of_vars}:

\begin{equation}\label{eq:nf_loss}
    \begin{aligned}
        \log p_\Xi(\bm{\xi}) &= \log p_Z(T^{-1}(\bm{\xi})) + \log \left| \det \left( J_{T^{-1}}(\bm{\xi}) \right) \right|  \\
    &= \log p_Z(\mathbf{z}_0) + \sum_{k=1}^{K} \log \left| \det \left( J_{T_k^{-1}}(\mathbf{z}_k) \right) \right|
    \end{aligned}
\end{equation}

The training loss is then the averaged log-likelihood over the dataset. 

\remark{The connection between copula-based mappings and normalising flows is worth noting. Under a Gaussian copula assumption, the Nataf transform in \eqref{eq:nataf} defines an explicit isoprobabilistic mapping between the physical inputs $\bm{\xi}$ and independent standard Gaussian latents $\mathbf{z}$, i.e., $\mathbf{z}=S_{\mathrm{nat}}(\bm{\xi})$ and $\bm{\xi}=T_{\mathrm{nat}}(\mathbf{z})$. This can be interpreted as a highly constrained, fixed-architecture ``flow-like'' construction once a copula family (and its parameters) is specified. Our approach generalizes this idea by learning a flexible transport map from data,
$\bm{\xi}=T_{\theta}(\mathbf{z})=T_{\theta,K}\circ\cdots\circ T_{\theta,1}(\mathbf{z})$,
where each layer is chosen to admit tractable evaluation of the forward map. This enables expressive dependence modelling beyond pre-specified copula forms while retaining compatibility with PCE.}

\subsection{Flow-Based PCE Scheme}
When used as a generative model, a trained normalising flow maps a simple base distribution (e.g., standard normal)
to a complex target distribution via an invertible transformation $\bm{\xi}=T(\mathbf{Z})$.
Given its bijective nature, the inverse map $T^{-1}$ sends data back to the base space, which is the direction
used for likelihood training in \eqref{eq:nf_loss}. Importantly, the map required for PCE construction is the
generative direction $T$ (from latent to physical space): it enables drawing i.i.d.\ samples
$\mathbf{Z}\sim p_Z$ and pushing them forward to obtain $\bm{\xi}=T(\mathbf{Z})$ for building the regression
design and evaluating the simulator. This observation motivates a novel scheme for constructing PCEs for problems
with complex, dependent inputs: we first learn a normalising flow to model the joint distribution of input data
$\bm{\xi}$, and then use the learned generative map $T$ directly within the PCE workflow.

To find the coefficients $\mathbf{h}$, we modify the standard least-squares problem from \eqref{eq:least_squares_problem} as
\begin{equation}
    \hat{\mathbf{h}} = \arg\min_{\mathbf{h} \in \mathbb{R}^P} \mathbb{E}\left[ \left(\mathcal{M}(T(\mathbf{Z})) - \mathbf{h}^\top \bm{\Psi}(\mathbf{Z})\right)^2 \right] + \lambda R(\mathbf{h})
    \label{eq:least_squares_regularized}
\end{equation}
A similar empirical setting in \eqref{eq:least_squares_problem} is also used. The complete data-driven workflow is presented in Algorithm~\ref{alg:flow_pce_formal}.

\begin{algorithm}[!t]
\caption{Flow-based PCE}
\label{alg:flow_pce_formal}
\begin{algorithmic}[1]
\Require Dataset of inputs $\mathcal{D}_\xi = \{\bm{\xi}^{(i)}\}_{i=1}^{N_\xi}$, computational model $\mathcal{M}(\cdot)$, total polynomial degree $p$, number of regression samples $N_s$, regularisation strength $\lambda$.
\Ensure PCE surrogate $\mathcal{M}_{\mathrm{PCE}}$.

\State \textbf{Train flow:} Fit a normalising flow $T_\theta$ on $\mathcal{D}_\xi$ via maximum likelihood, obtaining an invertible transport map $\bm{\xi}=T_{\theta}(\mathbf{z})$ from a chosen base distribution $p_Z$ to the data distribution.
\State \textbf{Sample and map:} Draw $N_s$ i.i.d. latent samples $\{\mathbf{z}^{(j)}\}_{j=1}^{N_s}\sim p_Z$ and map to physical space via $\bm{\xi}^{(j)} \leftarrow T_{\theta}(\mathbf{z}^{(j)})$.
\State \textbf{Evaluate model:} Compute $y^{(j)} \leftarrow \mathcal{M}(\bm{\xi}^{(j)})$ for $j=1,\ldots,N_s$.
\State \textbf{Construct basis:} Build an orthonormal polynomial basis $\{\Psi_{\bm{\alpha}}\}_{\bm{\alpha}\in\mathcal{A}}$ on latent space with total degree $|\bm{\alpha}|\leq p$ (e.g., Hermite polynomials for $p_Z=\mathcal{N}(\mathbf{0},\mathbf{I})$).
\State \textbf{Fit PCE:} Solve the regularised least-squares problem in \eqref{eq:least_squares_regularized} using the collected pairs $(\mathbf{z}^{(j)}, y^{(j)})$ to estimate $\hat{\mathbf{h}}$.
\State \textbf{Assemble surrogate:} Define $\mathcal{M}_{\mathrm{PCE}}(\mathbf{z}) = \sum_{\bm{\alpha}} \hat{h}_{\bm{\alpha}} \Psi_{\bm{\alpha}}(\mathbf{z})$.
\State \textbf{Post-process:} Compute desired statistics (e.g.\ mean, variance) of the output using the closed-form expressions in \eqref{eq:pce_statistic}.
\Return $\mathcal{M}_{\mathrm{PCE}}$.
\end{algorithmic}
\end{algorithm}

% In our case studies, we will compare different normalising flow architectures to evaluate their impact on the accuracy and efficiency of the PCE surrogate.

\subsection{Evaluation Metrics}

To evaluate the performance of the different UQ schemes, we employ three sets of metrics, each addressing a distinct aspect of the modeling workflow. The motivations for the metric designs are provided in Section~\ref{sec:theory}.

\subsubsection{Surrogate model accuracy} The \emph{normalized integrated root mean square error} (NIRMSE) is used to quantify how well the PCE surrogate reproduces the true statistics of the response. For a time-series response, denote by $\mathbf{\mu}_{\mathrm{ref}}$ and $\hat{\mathbf{\mu}}$ the reference and surrogate mean trajectories, and by $\mathbf{\sigma}_{\mathrm{ref}}$ and $\hat{\mathbf{\sigma}}$ the corresponding standard deviation trajectories. The NIRMSE is defined by
\begin{equation}
    \mathrm{NIRMSE} = \sqrt{ \Bigl(\tfrac{\|\mathbf{\mu}_{\mathrm{ref}} - \hat{\mathbf{\mu}}\|_2}{\|\mathbf{\mu}_{\mathrm{ref}}\|_2}\Bigr)^2 + \Bigl(\tfrac{\|\mathbf{\sigma}_{\mathrm{ref}} - \hat{\mathbf{\sigma}}\|_2}{\|\mathbf{\sigma}_{\mathrm{ref}}\|_2}\Bigr)^2 }.
\end{equation}
This single scalar summarises deviations in both mean and spread.

\subsubsection{Input distribution fidelity} When comparing the learned input distributions to the reference data distribution, we use different measures depending on whether the model permits exact density evaluation. For models like normalising flows, which yield an explicit density, we compute the forward Kullback–Leibler divergence $D_{\mathrm{KL}}(p_{\mathrm{data}}\|p_\theta)$ via Monte Carlo. When only samples are available (as in the copula fit), we resort to the Wasserstein distance, estimated from paired samples. 

\subsubsection{Transformation map smoothness} Finally, to assess the smoothness of the learned map $T$, this paper proposes a new \emph{Map Smoothness Index} (MSI). It is defined as the root-mean-square Frobenius norm of the Jacobian of $T$ over the latent distribution:
\begin{equation}
   \mathrm{MSI}(T;p_Z) = \sqrt{\frac{1}{M}\,\mathbb{E}_{\mathbf{z}\sim p_Z}\!\bigl[\|J_T(\mathbf{z})\|_F^2\bigr]}.
   \label{eq:msi}
\end{equation}
A smaller MSI indicates a less distorted, smoother map, which empirically correlates with improved PCE accuracy.

\subsection{Theoretical Error Analysis and Guidance}\label{sec:theory}

We derive a guidance-oriented upper bound for the generalisation error of the proposed Flow-based PCE framework, validate the proposed three metrics, and finally motivate the relationship between the MSI and the surrogate’s convergence rate.  

To start, it is assumed that the total mean-squared error \(\mathcal{E}_{\mathrm{total}}=\lVert Y-Y_{\mathrm{PCE}}\rVert_{L^2}\) can be decomposed into three components via the triangle inequality:
\begin{equation}
\mathcal{E}_{\mathrm{total}} \le \underbrace{\mathcal{E}_{\mathrm{flow}}}_{\text{distribution error}} + \underbrace{\mathcal{E}_{\mathrm{trunc}}}_{\text{approximation error}} + \underbrace{\mathcal{E}_{\mathrm{est}}}_{\text{estimation error}}
\end{equation}
and the computational model $\mathcal{M}(\xi)$ is locally Lipschitz continuous with constant \(L_{\mathcal{M}}\) on the support of the input distribution.  This holds for power-system DAEs over finite time intervals absent of bifurcations. Additionally, the normalising flow \(T_\theta(\mathbf{z})\) is assumed to be a continuously differentiable (\(C^1\)) diffeomorphism, a property satisfied by the rational quadratic splines considered in \cite{durkan2019neural}.

\subsubsection{Flow error \(\mathcal{E}_{\mathrm{flow}}\).}
This term captures the discrepancy between the true input distribution \(p_{\text{data}}\) and the learned flow density \(p_\theta\). Since \(\mathcal{M}\) is \(L_{\mathcal{M}}\)-Lipschitz,
applying the Lipschitz bound under the optimal \(W_2\) coupling yields
\begin{equation}
\mathcal{E}_{\mathrm{flow}} = \lVert \mathcal{M}(\mathbf{\Xi}) - \mathcal{M}(T_{\theta}(\mathbf{Z}))\rVert_{L^2} \;\leq \; L_{\mathcal{M}}\, W_2(p_{\text{data}},p_\theta)
\end{equation}
where \(W_2\) denotes the 2-Wasserstein distance.  Minimising the divergence between distributions during flow training typically reduces this error~\cite{villani2009optimal}.

\subsubsection{Truncation error \(\mathcal{E}_{\mathrm{trunc}}\).}
This error arises when projecting the composite function \(g(\mathbf{z})= \mathcal{M}(T_\theta(\mathbf{z}))\) onto a finite polynomial basis of degree \(p\).  From approximation theory for generalised polynomial chaos~\cite{xiu_numerical_2010}, if \(g\in H^1_{\mu_Z}\) (weighted Sobolev space) then
\begin{equation}\label{eq:first_trunc}
\mathcal{E}_{\mathrm{trunc}}\le C_1\,p^{-1}\,\lVert\nabla g(\mathbf{z})\rVert_{L^2(\mu_Z)}
\end{equation}
To make the dependence on the Map Smoothness Index (MSI) explicit, we can apply the chain rule, \(\nabla g(\mathbf{z}) = \nabla \mathcal{M}(T(\mathbf{z})) J_T(\mathbf{z})\), and separate the contribution of the physical model \(\mathcal{M}\) from the transport map \(T\):
\begin{equation}
    \lVert \nabla g(\mathbf{z}) \rVert_{L^2} \le \underbrace{\sup_{\mathbf{x}} \lVert \nabla \mathcal{M}(\mathbf{x}) \rVert}_{L_{\mathcal{M}}} \cdot \underbrace{\left( \mathbb{E}_Z [\lVert J_T(\mathbf{z}) \rVert_F^2] \right)^{1/2}}_{\mathrm{MSI}}
\end{equation}
Here, \(L_{\mathcal{M}}\) represents the maximum sensitivity of the physical model, and the second term is precisely the MSI. Substituting this back into \eqref{eq:first_trunc} yields the final bound:
\begin{equation}
    \mathcal{E}_{\mathrm{trunc}} \le \frac{C_1 \, L_{\mathcal{M}}}{p} \cdot \mathrm{MSI}
\end{equation}
This derivation clarifies that minimizing the MSI directly minimizes the upper bound of the truncation error.

\subsubsection{Estimation error \(\mathcal{E}_{\mathrm{est}}\).}
The final term accounts for the error in estimating the PCE coefficients using a finite sample.  When coefficients \(\hat{h}\) are estimated via \(\ell_1\)-regularised regression (LASSO) with \(N_s\) samples, high-dimensional statistical results~\cite{buhlmann2011statistics} give, with high probability,
\begin{equation}
\mathcal{E}_{\mathrm{est}} \le C_2\,\sqrt{\frac{s_0\log P}{N_s}}
\end{equation}
where \(s_0\) is the sparsity of the true expansion and \(P=\binom{M+p}{p}\) is the size of the candidate basis.  Smoother maps \(T\) often produce sparser expansions (smaller \(s_0\)), thereby further reducing this error.

\subsubsection{Summary and guidance.}
Combining the three components yields the overall bound
\begin{equation}
\mathcal{E}_{\mathrm{total}} \;\leq\; L_\mathcal{M}\,W_2(p_{\text{data}},p_\theta) + \frac{C_1 \, L_{\mathcal{M}}}{p}\mathrm{MSI}(T_{\theta}) + C_2 \sqrt{\frac{s_0\log P}{N_s}}
\end{equation}
which suggests selecting flow architectures that jointly achieve (i) good distribution fit  and (ii) moderate map sensitivity (low MSI), noting the empirical sweet spot (illustrated in Fig.~\ref{fig:bins_vs_error}): overly smooth/underfit maps can sacrifice the accuracy for the input, while overly complex maps may increase MSI and degrade PCE accuracy. If MSI remains high, increase the polynomial degree \(p\) (and/or simplify/regularize the flow), if MSI is low yet errors persist, increase the number of training samples \(N_s\) and/or adjust the regression regularization.

\section{Case Studies}\label{sec:case_study}

\subsection{Simulation Setup}

The test case is a modified IEEE 14-bus system, with its dynamic behaviour simulated using the ANDES framework \cite{cui_hybrid_2021}. The system includes five generators: four synchronous generators (SGs) and one solar PV plant. The input uncertainty is introduced through the active power injections from three of these generators, which are treated as a 3-dimensional random vector $\bm{\xi}$. We will use different synthetic datasets for $\bm{\xi}$ to test the methods under various conditions. Note that the uncertainty in the initial operating point stems from complex forecasting errors, resulting in an input distribution with features like nonlinear dependencies and multimodality. As standard parametric models cannot capture this complexity, an advanced, data-driven generative model is essential to reliably represent the true uncertainty from observational data.

The synchronous generators are modelled using the standard GENROU model, a TGOV1 turbine model, and an EXST1-type static excitation system. The renewable generator is represented by standard models for the generator, exciter, and plant-level controls using REGCA1, REECA1 and REPCA1 models. To introduce significant non-linearity into the model response and create a non-trivial uncertainty propagation problem, we simulate two line trip events during the dynamic simulation. 
% Without such perturbations, the problem would largely degrade to a simpler study of random initial conditions determined by the power flow.

% \remark{Uncertainty in the initial operating point stems from complex forecasting errors, resulting in an input distribution with features like nonlinear dependencies and multimodality. As standard parametric models cannot capture this complexity, an advanced, data-driven generative model is essential to reliably represent the true uncertainty from observational data.}

\subsection{Test on Gaussian Copula Data}

In this first test, we evaluate the performance of the methods on synthetic data generated from a distribution with a \emph{known} dependence structure. This serves as a baseline to see how the general-purpose flow-based method compares to the classical approach when the data perfectly matches the assumptions of the classical method.

The 3-dimensional random input vector $\bm{\xi} = (\xi_1, \xi_2, \xi_3)$ is generated from a joint distribution defined by a Gaussian copula with the following marginal distributions:

\begin{itemize}
    \item $\xi_1 \sim \text{Beta}(2, 5)$ on $[0, 1]$
    \item $\xi_2 \sim \text{Uniform}(0.2, 0.8)$
    \item $\xi_3 \sim \text{TruncNormal}(\mu=0.5, \sigma=0.1)$ on $[0, 1]$
\end{itemize}

The dependence is governed by a Gaussian copula with the correlation matrix $\mathbf{R}_0$:

\begin{equation}
    \mathbf{R}_0 = 
    \begin{pmatrix}
        1.0 & 0.5 & 0.2 \\
        0.5 & 1.0 & 0.3 \\
        0.2 & 0.3 & 1.0
    \end{pmatrix}
\end{equation}

\begin{figure}[!t]
    \centering
    \includegraphics[width=0.80\columnwidth]{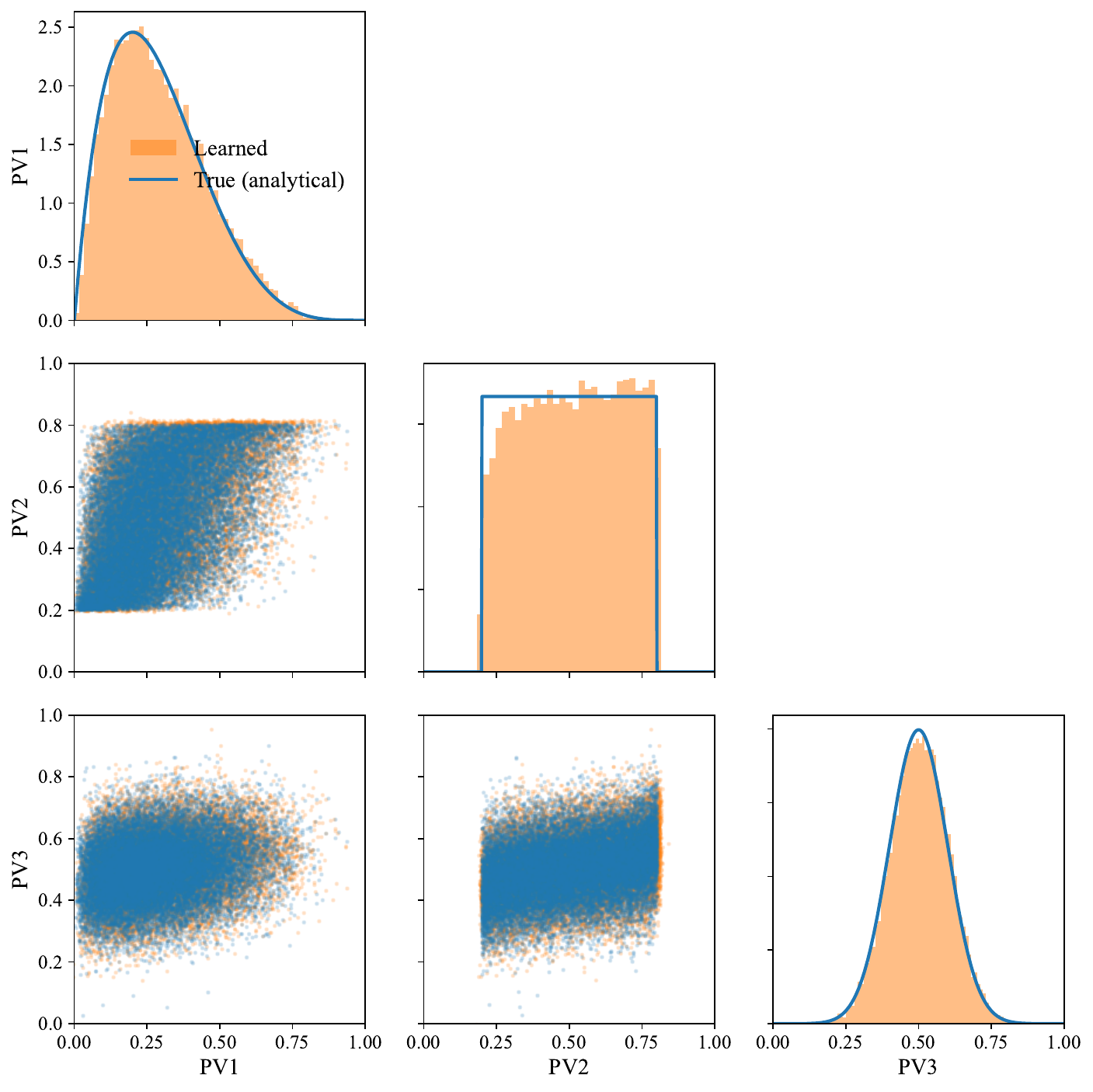} % Replace with actual figure
    \caption{Pair plot visualizing the joint distribution learned by the spline-based normalising flow on the Gaussian copula dataset.}
    \label{fig:dist_flow_only}
\end{figure}

\begin{figure}[!t]
    \centering
    
    % --- First Subfigure ---
    \begin{subfigure}{\columnwidth}
        \centering % Essential to center the content within the subfigure
        \hspace{3mm} % <--- Adjust this value to shift the image right
        \includegraphics[width=0.8\linewidth]{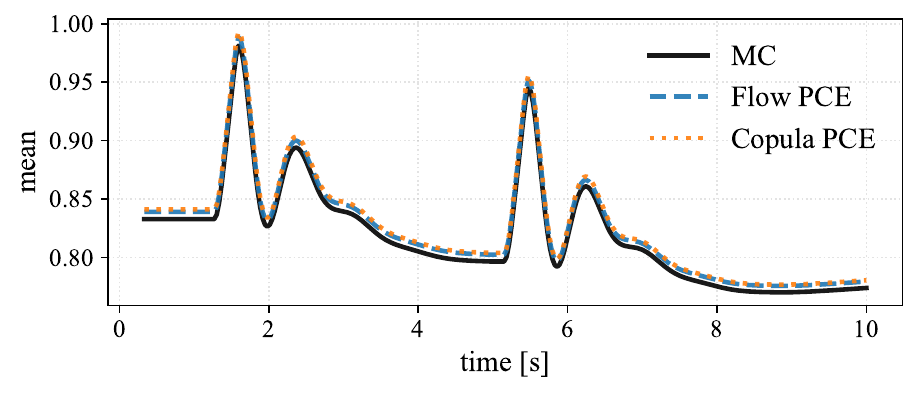}
        \caption{Mean trajectory comparison on the rotor angle.}
        \label{fig:mean_copula}
    \end{subfigure}
    
    \vfill
    
    % --- Second Subfigure ---
    \begin{subfigure}{\columnwidth}
        \centering 
        \hspace{3mm} % <--- Same shift here
        \includegraphics[width=0.8\linewidth]{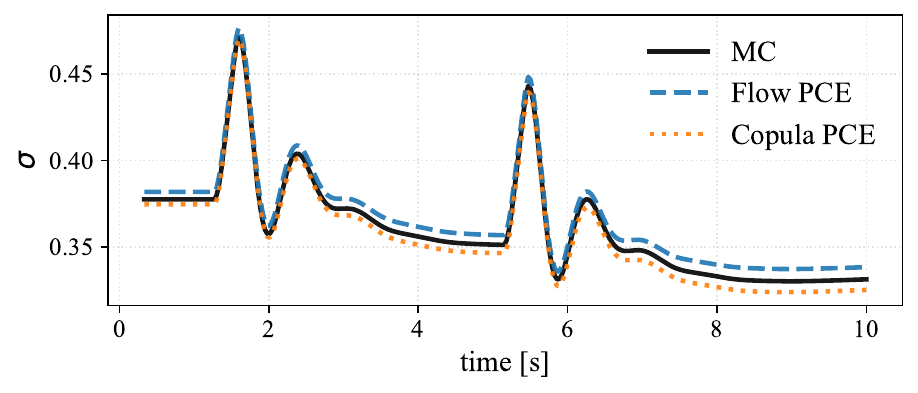}
        \caption{Standard deviation trajectory comparison on the rotor angle.}
        \label{fig:std_copula}
    \end{subfigure}

    \caption{Comparison of the PCE surrogate outputs for the Gaussian copula test case.}
    \label{fig:results_copula_data}
\end{figure}

We learn the input distribution from a dataset of 100,000 samples using both a Gaussian copula model and a spline-based normalising flow. Figure~\ref{fig:dist_flow_only} visualizes the joint distribution learned by the normalising flow. Subsequently, we use these learned transformations to construct PCE surrogates according to Algorithm~\ref{alg:flow_pce_formal}. The statistical outputs of the resulting surrogates are shown in Figure~\ref{fig:results_copula_data}, and the quantitative performance is summarized in Table~\ref{tab:results_copula}.

\begin{table}[!t]
\footnotesize                      % one step smaller than normalsize
\setlength{\tabcolsep}{3pt}          % tighten column spacing
\caption{Quantitative comparison of surrogate‑model accuracy
        on the Gaussian-copula dataset.}
\label{tab:results_copula}
\centering
\begin{tabularx}{\columnwidth}{l *{4}{>{\centering\arraybackslash}X}}
\toprule
\textbf{Method} & \textbf{Wasserstein} & \textbf{Mean} & \textbf{Std.\ dev.} & \textbf{NIRMSE}\\
                & \textbf{dist.}      & \textbf{error}& \textbf{error}       &                \\
\midrule
Copula & \(2.04 \times 10^{-3}\) & \(1.86 \times 10^{-2}\) & \(1.59 \times 10^{-2}\) & \(2.45 \times 10^{-2}\)\\
Flow   & \(3.90 \times 10^{-3}\) & \(8.02 \times 10^{-3}\) & \(2.54 \times 10^{-2}\) & \(2.66 \times 10^{-2}\)\\
\bottomrule
\end{tabularx}
\end{table}

As shown in Table~\ref{tab:results_copula}, both methods achieve a similar level of accuracy, with the copula-based approach having a slight edge in overall distribution fidelity and the NIRMSE score. However, it is crucial to recognize that this test case is intentionally designed to \emph{favour} the classical Copula-PCE method. By generating data from a known Gaussian copula, we provide the algorithm with perfect prior knowledge of the underlying dependence structure. Despite this significant handicap, the normalising flow-based PCE, which starts with no such prior assumptions and learns the entire joint distribution from data alone, achieves a remarkably comparable level of performance.

\subsection{Test on Multimodal Data}

The true value of the proposed Flow-based PCE framework shines when the input distribution cannot be accurately described by a simple model like a Gaussian copula. Forcing a classical copula model onto such data leads to a poor characterization of the source uncertainty, which inevitably degrades the accuracy of the resulting PCE surrogate. To demonstrate this, we now test the methods on a synthetic dataset drawn from a bimodal distribution. The input vector $\bm{\xi}$ is generated from a truncated Gaussian mixture distribution. The base distribution is an equal-weighted mixture of two 3-dimensional normal distributions, $p_{mix}(\bm{\xi}) = 0.5 \cdot \mathcal{N}(\bm{\xi}; \bm{\mu}_1, \mathbf{\Sigma}) + 0.5 \cdot \mathcal{N}(\bm{\xi}; \bm{\mu}_2, \mathbf{\Sigma})$, with parameters:
\begin{align}
    \bm{\mu}_1 &= [0.25, 0.30, 0.35]^\top \\
    \bm{\mu}_2 &= [0.75, 0.70, 0.65]^\top \\
    \mathbf{\Sigma} &= 0.02 \cdot \mathbf{I}_3
\end{align}
The final distribution $p_\Xi$ is obtained by truncating this mixture distribution to the unit hypercube $[0, 1]^3$.

\begin{figure*}[!t]
    \centering
    \captionsetup[subfigure]{font=footnotesize} % <— local tweak
    \subfloat[Gaussian Copula Fit]{\includegraphics[width=0.37\textwidth]{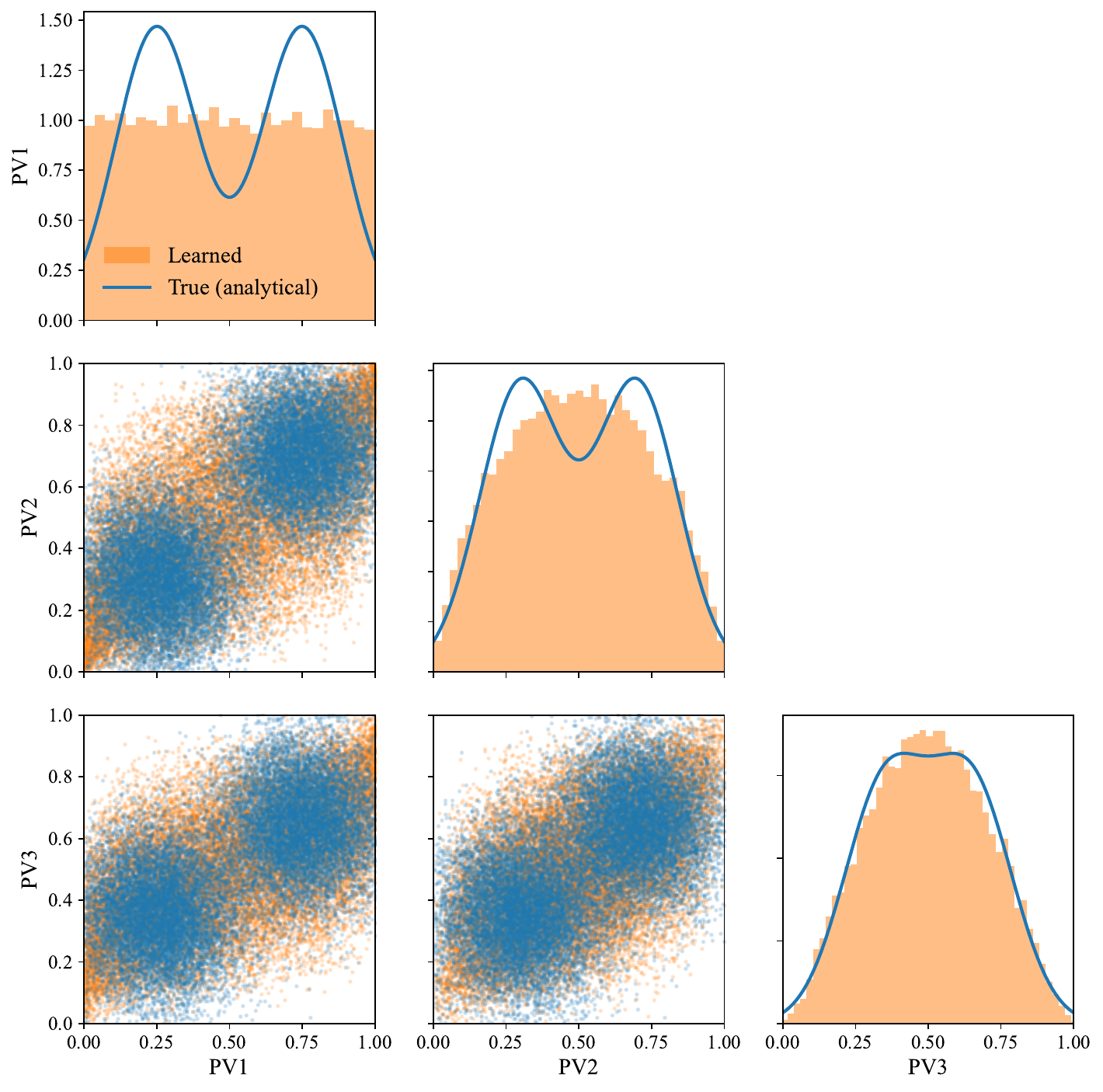}\label{fig:dist_copula_multimodal}}
    \hspace{0.05\textwidth}%  ← fixed, small gap
    \subfloat[Normalising Flow Fit]{\includegraphics[width=0.37\textwidth]{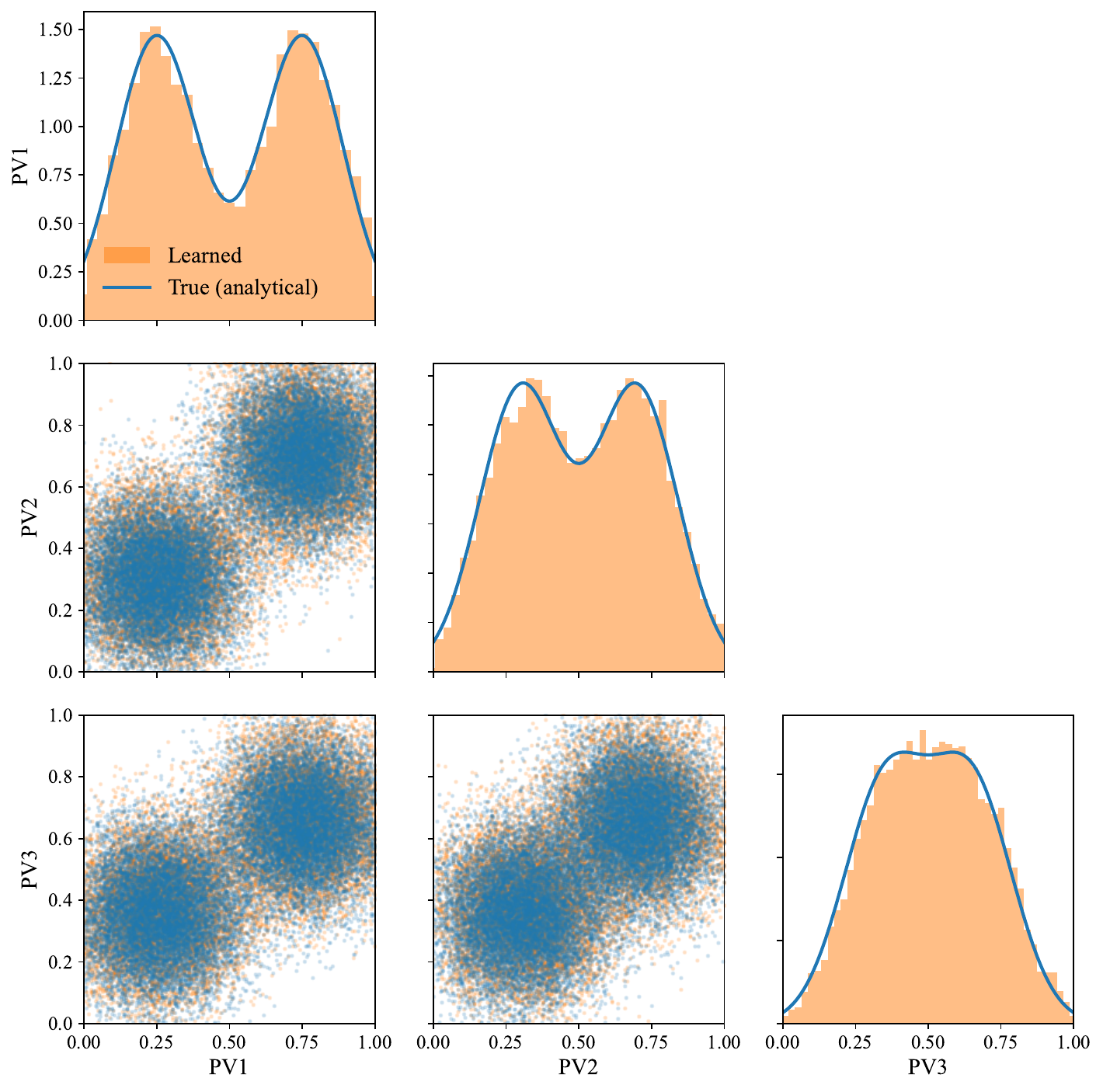}\label{fig:dist_flow_multimodal}}
    \caption{Comparison of learned distributions on the bimodal dataset. The classical copula model (a) fails to capture the two distinct modes of the data, representing it as a single unimodal distribution. The normalising flow model (b) successfully learns the bimodal structure.}
    \label{fig:dist_multimodal_data}
\end{figure*}

\begin{figure}[!t]
    \centering
    \begin{subfigure}{\columnwidth}
        \includegraphics[width=0.8\linewidth]{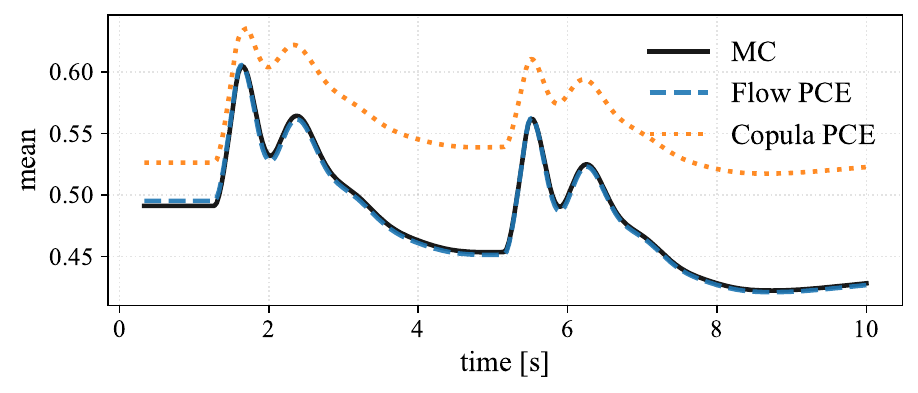}
        \caption{Mean trajectory comparison.}
    \end{subfigure}

    \vspace{0.5em}

    \begin{subfigure}{\columnwidth}
        \includegraphics[width=0.8\linewidth]{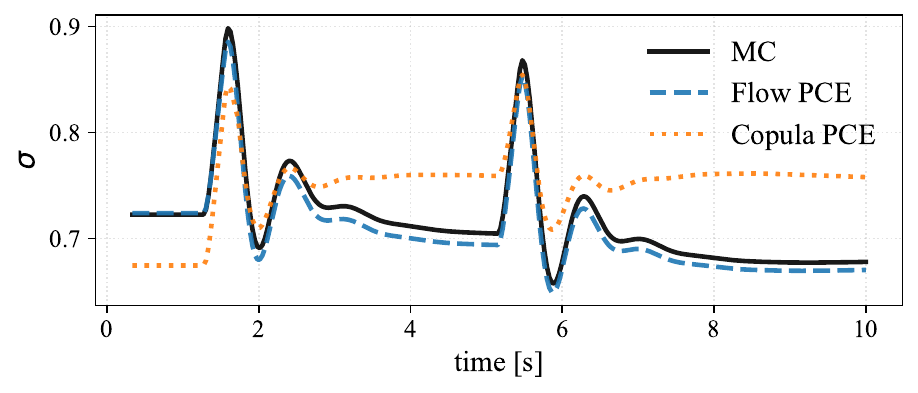}
        \caption{Standard deviation trajectory comparison.}
    \end{subfigure}
    \caption{Comparison of PCE surrogate outputs for the multimodal test case.}
    \label{fig:results_multimodal_data}
\end{figure}

We generate a 30,000-sample dataset to train both a Gaussian copula and a 5-layer Neural Spline Flow (NSF) which uses piecewise rational quadratic splines to model smooth monotonic transformations \cite{durkan_spline_2019}. For a fair comparison, PCE surrogates are then constructed for each method using identical settings: a polynomial order of 4 and LASSO ($\ell_1$) regularized regression. As shown in Fig.~\ref{fig:dist_multimodal_data}, the classical Gaussian copula model fails to capture the bimodal nature of the data, fitting a single elliptical distribution instead. In contrast, the normalising flow successfully represents the two distinct modes. 
% This difference in distribution fidelity has a direct impact on the PCE surrogate's accuracy. 
Fig.~\ref{fig:results_multimodal_data} clearly shows that the surrogate built using the poorly-fitted copula deviates significantly from the Monte Carlo reference, while the Flow-based PCE surrogate remains highly accurate. The quantitative metrics in Table~\ref{tab:results_multimodal} confirm this observation, with the Flow-based PCE achieving an error that is more than three times lower than that of the Copula-PCE.

\begin{table}[!t]
\centering
\footnotesize                      % one step smaller than normalsize
\setlength{\tabcolsep}{3pt}          % tighten column spacing
\caption{Quantitative comparison of surrogate-model accuracy on the multimodal dataset.}
\label{tab:results_multimodal}
\renewcommand{\arraystretch}{1.3} % Increase row spacing for readability
\begin{tabular}{lcccc}
\toprule
\bfseries Method & \begin{tabular}{@{}c@{}} \bfseries Wasserstein \\\bfseries Dist. \end{tabular} & \begin{tabular}{@{}c@{}} \bfseries Mean \\\bfseries Error \end{tabular} & \begin{tabular}{@{}c@{}} \bfseries Std. Dev. \\\bfseries Error \end{tabular} & \bfseries NIRMSE \\
\midrule
Copula & $1.23 \times 10^{-2}$ & $9.24 \times 10^{-2}$ & $1.52 \times 10^{-1}$ & $1.78 \times 10^{-1}$ \\
Flow   & $1.01 \times 10^{-2}$ & $7.05 \times 10^{-3}$ & $8.00 \times 10^{-3}$ & $1.07 \times 10^{-2}$ \\
\bottomrule
\end{tabular}
\end{table}

\subsection{On the Selection of the Normalising Flow}
\begin{figure}[!t]
    \centering
    \includegraphics[width=0.75\columnwidth]{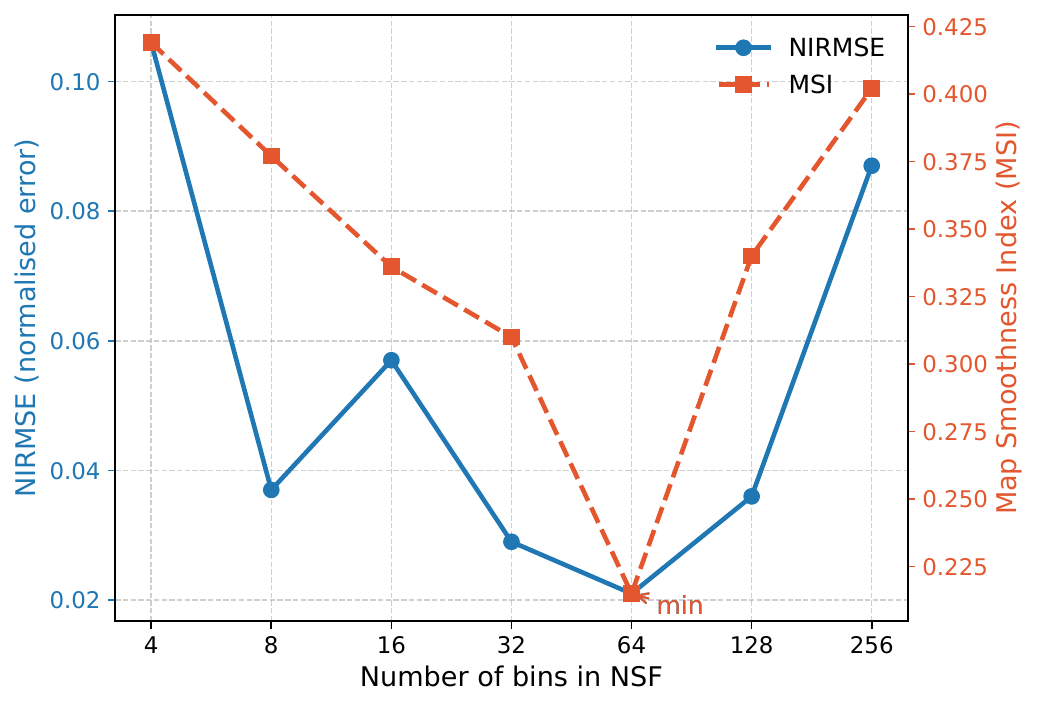} % Replace with your 'bins.png'
    \caption{NIRMSE vs. MSI as the number of NSF bins is varied. The optimal performance is achieved at a "sweet spot" of smoothness, avoiding both underfitting and overfitting of the transformation map.}
    \label{fig:bins_vs_error}
\end{figure}

This section investigates the impact of different types of flow, specifically, its architecture and smoothness, on the accuracy of the resulting PCE surrogate. We use the multimodal Gaussian mixture distribution from the previous section as our testbed, as its complexity provides a challenging benchmark. We compare the previously used Neural Spline Flow (NSF) with two additional flow architectures,
namely Masked Autoregressive Flow (MAF) and NICE (a coupling-layer flow). These architectures differ in how they parameterise the bijective map.  MAF \cite{papamakarios_maf_2017} factorises the joint density into a product of conditionals using autoregressive transformations parameterised by neural networks. NICE \cite{dinh_nice_2015} employs simple additive coupling layers that alternate between dimensions. While all three architectures are highly expressive and achieve similar KL divergences when trained on the data, their underlying structures produce transformation maps of varying smoothness. NSF uses piecewise rational quadratic splines, which are inherently smooth. In contrast, MAF and NICE rely on affine transformations conditioned by complex neural networks, which can learn less smooth, more ``jagged'' mappings. The results in Table~\ref{tab:results_flow_comparison} verify the idea: the spline flow, the smoothest flow, yields the most accurate PCE surrogate. We also test the effect of varying the number of spline bins within the NSF architecture. As shown in Fig.~\ref{fig:bins_vs_error}, an optimal ``sweet spot'' exists where the map is expressive enough to fit the data yet smooth enough to keep the MSI low.

\begin{table}[!t]
\centering
\caption{Impact of flow architecture on smoothness and PCE accuracy.}
\label{tab:results_flow_comparison}
\renewcommand{\arraystretch}{1.3}
\begin{tabular}{lcc}
\toprule
\bfseries Flow Architecture & \bfseries MSI & \bfseries NIRMSE \\
\midrule
NSF (Spline)    & 0.215 & 0.021 \\
MAF             & 0.273 & 0.103 \\
NICE            & 0.241 & 0.086 \\
\bottomrule
\end{tabular}
\end{table}

This study demonstrates that the proposed MSI can be used as a practical tool for model selection. For a fixed dataset size, one can tune the flow's hyperparameters to find the smoothest possible map that still accurately represents the data, thereby optimizing the conditions for building a high-fidelity PCE surrogate.

\subsection{Case Study for GB Transmission Network}
 We illustrate scalability on a Great Britain transmission test system with 2224 buses using synthetic multimodal renewable-injection data motivated by weather-regime switching. We vary the input dimension by selecting $M\!\in\!\{6,12,18\}$ aggregated renewable sites, fit either a normalising flow or a Gaussian-copula model to the joint injections, and then construct the corresponding PCE surrogate with various basis sizes. Overall, Flow-based PCE maintains lower Wasserstein and lower NIRMSE than Copula--PCE as the dimension increases. Table~\ref{tab:results_gb} also reports model-training time, highlighting the additional offline cost of expressive flows.

\begin{table}[!t]
\centering
\scriptsize
\setlength{\tabcolsep}{2.4pt}
\renewcommand{\arraystretch}{1.05}
\caption{GB scalability results for $M\in\{6,12,18\}$ uncertain injections.}
\label{tab:results_gb}
\begin{tabular}{cc l cc cc}
\toprule
\textbf{Dim.} & \textbf{Basis Size} & \textbf{Method} & \textbf{Params} & \textbf{Train (s)} & \textbf{Wasserstein} & \textbf{NIRMSE} \\
\midrule
\multirow{2}{*}{6}  & \multirow{2}{*}{210}  & Flow   & 269k & 67.6  & $2.57\times 10^{-2}$ & $1.31\times 10^{-2}$ \\
                    &                       & Copula & 15   & 15.4  & $3.64\times 10^{-1}$ & $8.40\times 10^{-2}$ \\
\midrule
\multirow{2}{*}{12} & \multirow{2}{*}{455}  & Flow   & 455k & 127.8 & $3.91\times 10^{-2}$ & $1.54\times 10^{-2}$ \\
                    &                       & Copula & 66   & 62.2  & $3.09\times 10^{-1}$ & $5.58\times 10^{-2}$ \\
\midrule
\multirow{2}{*}{18} & \multirow{2}{*}{1330} & Flow   & 640k & 124.2 & $5.35\times 10^{-2}$ & $1.05\times 10^{-1}$ \\
                    &                       & Copula & 153  & 81.7  & $3.30\times 10^{-1}$ & $1.47\times 10^{-1}$ \\
\bottomrule
\end{tabular}
\end{table}

\section{Conclusion}\label{sec:conclusion}

This paper proposes Flow-based PCE, a unified uncertainty quantification framework that jointly addresses both input modelling error and uncertainty propagation error. The proposed method employs a normalising flow to learn an expressive and invertible transport map for complex correlated input distributions, including multimodal distributions, and then exploits this map to construct a classical PCE surrogate for efficient uncertainty propagation. The main advantage of the framework is that it combines the representational flexibility of modern generative models with the theoretical foundation of PCE. Moreover, we introduce the Map Smoothness Index (MSI) and derive an explicit error bound that relates the quality of the learned transport map to PCE convergence, thereby offering theoretical and practical guidance beyond what is typically available from purely data-driven surrogates. Simulation results demonstrate that the proposed method consistently outperforms state-of-the-art PCE methods in terms of modelling accuracy and scalability.

% This paper proposed Flow-based PCE, a unified uncertainty-quantification framework that \emph{jointly} addresses (i) \emph{input modelling error} and (ii) \emph{uncertainty propagation error}. The proposed method learns an expressive, invertible transport map with a normalising flow to represent complex, correlated (including multimodal) input distributions, and then reuse this map to enable a classical PCE surrogate for fast propagation. The key strength is that this combination preserves the flexibility of state-of-the-art generative models while retaining the theoretical structure of PCE. We introduce the Map Smoothness Index (MSI) and provide an explicit bound that links map quality to PCE convergence, offering guidance that purely data-driven surrogates typically cannot provide. While the theoretical result is verified on large-scale power system dynamics, it can be applied to any expensive simulator where realistic input distributions must be learned from data and propagated efficiently.

{
\def\IEEEbibitemsep{00pt}
\scriptsize
\bibliographystyle{IEEEtran}
\bibliography{Refs.bib}

@article{buffelli2024exact,
  title={Exact, tractable gauss-newton optimization in deep reversible architectures reveal poor generalization},
  author={Buffelli, Davide and McGowan, Jamie and Xu, Wangkun and Cioba, Alexandru and Shiu, Da-shan and Hennequin, Guillaume and Bernacchia, Alberto},
  journal={Advances in Neural Information Processing Systems},
  volume={37},
  pages={133541--133570},
  year={2024}
}

@book{xiu_numerical_2010,
  location = {Princeton (N.J.)},
  title = {Numerical methods for stochastic computations: a spectral method approach},
  isbn = {978-0-691-14212-8},
  shorttitle = {Numerical methods for stochastic computations},
  publisher = {Princeton university press},
  author = {Xiu, Dongbin},
  year = {2010}
}

@article{Milanovic_ProbStab2017,
  author  = {Jovica V. Milanović},
  title   = {Probabilistic Stability Analysis: The Way Forward for Stability Analysis of Sustainable Power Systems},
  journal = {Philosophical Transactions of the Royal Society~A: Mathematical, Physical and Engineering Sciences},
  volume  = {375},
  number  = {2100},
  pages   = {20160296},
  year    = {2017},
}

@article{cui_hybrid_2021,
  author  = {Cui, Hantao and Li, Fangxing and Tomsovic, Kevin},
  journal = {IEEE Transactions on Power Systems},
  title   = {Hybrid Symbolic-Numeric Framework for Power System Modeling and Analysis},
  year    = {2021},
  volume  = {36},
  number  = {2},
  pages   = {1373-1384}
}

@article{jakeman_polynomial_2019,
  title   = {Polynomial chaos expansions for dependent random variables},
  volume  = {351},
  pages   = {643--666},
  journal = {Computer Methods in Applied Mechanics and Engineering},
  author  = {Jakeman, John and Franzelin, Fabian and Narayan, Akil and Eldred, Michael and Plfueger, Dirk},
  year    = {2019},
  month   = jul
}

@misc{durkan_spline_2019,
  title   = {Neural Spline Flows},
  url     = {http://arxiv.org/abs/1906.04032},
  number  = {arXiv:1906.04032},
  publisher = {arXiv},
  author  = {Durkan, Conor and Bekasov, Artur and Murray, Iain and Papamakarios, George},
  year    = {2019},
  month   = dec,
  note    = {arXiv:1906.04032 [stat]}
}

@misc{dinh_nice_2015,
  title   = {{NICE}: Non-linear Independent Components Estimation},
  url     = {http://arxiv.org/abs/1410.8516},
  shorttitle = {{NICE}},
  number  = {arXiv:1410.8516},
  publisher = {arXiv},
  author  = {Dinh, Laurent and Krueger, David and Bengio, Yoshua},
  year    = {2015},
  month   = apr,
  note    = {arXiv:1410.8516 [cs]}
}

@article{wu_probabilistic_1983,
  title   = {Probabilistic dynamic security assessment of power systems-I: Basic model},
  volume  = {30},
  pages   = {148--159},
  number  = {3},
  journal = {IEEE Transactions on Circuits and Systems},
  author  = {Wu, F. and Tsai, Yu-Kun},
  year    = {1983},
  month   = mar
}

@techreport{bird_integrating_2013,
  title   = {Integrating Variable Renewable Energy: Challenges and Solutions},
  shorttitle = {Integrating Variable Renewable Energy},
  pages   = {NREL/TP--6A20--60451, 1097911},
  number  = {NREL/TP--6A20-60451, 1097911},
  author  = {Bird, L. and Milligan, M. and Lew, D.},
  year    = {2013},
  month   = sep
}

@article{siano_demand_2014,
  title   = {Demand response and smart grids—A survey},
  volume  = {30},
  pages   = {461--478},
  journal = {Renewable and Sustainable Energy Reviews},
  author  = {Siano, Pierluigi},
  year    = {2014},
  month   = feb
}

@article{foley_current_2012,
  title   = {Current methods and advances in forecasting of wind power generation},
  volume  = {37},
  pages   = {1--8},
  number  = {1},
  journal = {Renewable Energy},
  author  = {Foley, Aoife M. and Leahy, Paul G. and Marvuglia, Antonino and {McKeogh}, Eamon J.},
  year    = {2012},
  month   = jan
}

@techreport{sudret_uncertainty_2007,
  author       = {Sudret, Bruno},
  title        = {Uncertainty Propagation and Sensitivity Analysis in Mechanical Models — Contributions to Structural Reliability and Stochastic Spectral Methods},
  type         = {Habilitation à diriger des recherches},
  institution  = {Université Blaise Pascal},
  address      = {Clermont-Ferrand, France},
  year         = {2007},
  pages        = {229}
}

@article{crestaux_polynomial_2009,
  title   = {Polynomial chaos expansion for sensitivity analysis},
  volume  = {94},
  pages   = {1161--1172},
  number  = {7},
  journal = {Reliability Engineering \& System Safety},
  author  = {Crestaux, Thierry and Le Maıˆtre, Olivier and Martinez, Jean-Marc},
  year    = {2009},
  month   = jul
}

@article{ernst_convergence_2012,
  title   = {On the convergence of generalized polynomial chaos expansions},
  volume  = {46},
  pages   = {317--339},
  number  = {2},
  journal = {ESAIM: Mathematical Modelling and Numerical Analysis},
  author  = {Ernst, Oliver G. and Mugler, Antje and Starkloff, Hans-Jörg and Ullmann, Elisabeth},
  year    = {2012},
  month   = mar
}

@thesis{lataniotis_data-driven_2019,
  title   = {Data-driven uncertainty quantification for high-dimensional engineering problems},
  institution = {{ETH} Zurich},
  type    = {phdthesis},
  author  = {Lataniotis, Christos},
  year    = {2019},
  month   = nov
}

@article{bond-taylor_deep_2022,
  title   = {Deep Generative Modelling: A Comparative Review of {VAEs}, {GANs}, Normalizing Flows, Energy-Based and Autoregressive Models},
  volume  = {44},
  pages   = {7327--7347},
  number  = {11},
  journal = {IEEE Transactions on Pattern Analysis and Machine Intelligence},
  author  = {Bond-Taylor, Sam and Leach, Adam and Long, Yang and Willcocks, Chris G.},
  year    = {2022},
  month   = nov
}

@article{sudret2008global,
  title   = {Global sensitivity analysis using polynomial chaos expansions},
  author  = {Sudret, Bruno},
  journal = {Reliability Engineering \& System Safety},
  volume  = {93},
  number  = {7},
  pages   = {964--979},
  year    = {2008}
}

@article{rahman_polynomial_2018,
  title   = {A polynomial chaos expansion in dependent random variables},
  volume  = {464},
  pages   = {749--775},
  number  = {1},
  journal = {Journal of Mathematical Analysis and Applications},
  author  = {Rahman, Sharif},
  year    = {2018},
  month   = aug
}

@article{torre_data-driven_2019,
  title   = {Data-driven polynomial chaos expansion for machine learning regression},
  volume  = {388},
  pages   = {601--623},
  journal = {Journal of Computational Physics},
  author  = {Torre, E. and Marelli, S. and Embrechts, P. and Sudret, B.},
  year    = {2019},
  month   = jul
}

@article{feinberg_multivariate_2018,
  title   = {Multivariate Polynomial Chaos Expansions with Dependent Variables},
  volume  = {40},
  pages   = {A199--A223},
  number  = {1},
  journal = {SIAM Journal on Scientific Computing},
  author  = {Feinberg, Jonathan and Eck, Vinzenz Gregor and Langtangen, Hans Petter},
  year    = {2018},
  month   = jan
}

@article{soize_physical_2004,
  title   = {Physical Systems with Random Uncertainties: Chaos Representations with Arbitrary Probability Measure},
  volume  = {26},
  pages   = {395--410},
  number  = {2},
  journal = {SIAM Journal on Scientific Computing},
  author  = {Soize, Christian and Ghanem, Roger},
  year    = {2004},
  month   = jan
}

@article{Ni_powerflow,
  author  = {Ni, Fei and Nguyen, Phuong H. and Cobben, Joseph F. G.},
  journal = {IEEE Transactions on Power Systems},
  title   = {Basis-Adaptive Sparse Polynomial Chaos Expansion for Probabilistic Power Flow},
  year    = {2017},
  volume  = {32},
  number  = {1},
  pages   = {694-704}
}

@misc{qiu_nonintrusive_2020,
  author={Qiu, Yiwei and Lin, Jin and Chen, Xiaoshuang and Liu, Feng and Song, Yonghua},
  journal={IEEE Transactions on Power Systems}, 
  title={Nonintrusive Uncertainty Quantification of Dynamic Power Systems Subject to Stochastic Excitations}, 
  year={2021},
  volume={36},
  number={1},
  pages={402-414},
}

@misc{metivier_efficient_2019,
title = {Efficient polynomial chaos expansion for uncertainty quantification in power systems},
journal = {Electric Power Systems Research},
volume = {189},
pages = {106791},
year = {2020},
issn = {0378-7796},
author = {David Métivier and Marc Vuffray and Sidhant Misra},
}

@article{xu_propagating_2019,
  title   = {Propagating Uncertainty in Power System Dynamic Simulations Using Polynomial Chaos},
  volume  = {34},
  pages   = {338--348},
  number  = {1},
  journal = {IEEE Transactions on Power Systems},
  author  = {Xu, Yijun and Mili, Lamine and Sandu, Adrian and Spakovsky, Michael R. von and Zhao, Junbo},
  year    = {2019},
  month   = jan
}

@inproceedings{ye_uncertainty_2021,
  title   = {Uncertainty Quantification of Loads and Correlated {PVs} on Power System Dynamic Simulations},
  eventtitle = {2021 {IEEE}/{IAS} Industrial and Commercial Power System Asia (I\&{CPS} Asia)},
  pages   = {44--49},
  booktitle = {2021 {IEEE}/{IAS} Industrial and Commercial Power System Asia (I\&{CPS} Asia)},
  author  = {Ye, Ketian and Zhao, Junbo and Duan, Nan and Maldonado, Daniel Adrian},
  year    = {2021},
  month   = jul
}

@article{fan_uncertainty_2021,
  title   = {Uncertainty Evaluation Algorithm in Power System Dynamic Analysis With Correlated Renewable Energy Sources},
  volume  = {36},
  pages   = {5602--5611},
  number  = {6},
  journal = {IEEE Transactions on Power Systems},
  author  = {Fan, Miao and Li, Zhengshuo and Ding, Tao and Huang, Lengcheng and Dong, Feng and Ren, Zhouyang and Liu, Chengxi},
  year    = {2021},
  month   = nov
}

@article{rosenblatt1952remarks,
  author  = {Rosenblatt, Murray},
  title   = {Remarks on a Multivariate Transformation},
  journal = {The Annals of Mathematical Statistics},
  year    = {1952},
  volume  = {23},
  number  = {3},
  pages   = {470--472},
  month   = sep,
  publisher = {Institute of Mathematical Statistics}
}

@article{noh_reliability-based_2009,
	title = {Reliability-based design optimization of problems with correlated input variables using a Gaussian Copula},
	volume = {38},
	pages = {1--16},
	number = {1},
	journal = {Structural and Multidisciplinary Optimization},
	shortjournal = {Struct Multidisc Optim},
	author = {Noh, Yoojeong and Choi, K. K. and Du, Liu},
	date = {2009-03},
	langid = {english},
}

@book{villani2009optimal,
  title={Optimal transport: old and new},
  author={Villani, C{\'e}dric},
  volume={338},
  year={2009},
  publisher={Springer},
  address={Berlin, Heidelberg}
}

@book{buhlmann2011statistics,
  title={Statistics for high-dimensional data: methods, theory and applications},
  author={B{\"u}hlmann, Peter and Van De Geer, Sara},
  year={2011},
  publisher={Springer Science \& Business Media}
}

@inproceedings{durkan2019neural,
  title={Neural spline flows},
  author={Durkan, Conor and Bekasov, Artur and Murray, Iain and Papamakarios, George},
  booktitle={Advances in neural information processing systems},
  volume={32},
  year={2019}
}

@inproceedings{papamakarios_maf_2017,
  author    = {Papamakarios, George and Pavlakou, Th{\'e}o and Murray, Iain},
  title     = {Masked Autoregressive Flow for Density Estimation},
  booktitle = {Advances in Neural Information Processing Systems},
  volume    = {30},
  year      = {2017}
}

@article{tan_gaussian_2026,
	title = {Gaussian processes in power systems: Techniques, applications, and future works},
	volume = {402},
	issn = {0306-2619},
	shorttitle = {Gaussian processes in power systems},
	journal = {Applied Energy},
	shortjournal = {Applied Energy},
	author = {Tan, Bendong and Su, Tong and Weng, Yu and Ye, Ketian and Pareek, Parikshit and Vorobev, Petr and Nguyen, Hung and Zhao, Junbo and Deka, Deepjyoti},
	urldate = {2025-12-10},
	date = {2026-01-01},
}

@article{fang2018spatial,
  title={Modelling wind power spatial-temporal correlation in multi-interval optimal power flow: A sparse correlation matrix approach},
  author={Fang, Xin and Hodge, Bri-Mathias and Du, Ershun and Zhang, Ning and Li, Fangxing},
  journal={Applied Energy},
  volume={230},
  pages={531--539},
  year={2018},
  publisher={Elsevier},
  doi={10.1016/j.apenergy.2018.08.123}
}

@article{schafer2018nonGaussian,
  title={Non-Gaussian power grid frequency fluctuations characterized by L{\'e}vy-stable laws and superstatistics},
  author={Sch{\"a}fer, Benjamin and Beck, Christian and Aihara, Kazuyuki and Witthaut, Dirk and Timme, Marc},
  journal={Nature Energy},
  volume={3},
  pages={119},
  year={2018},
  doi={10.1038/s41560-017-0058-z}
}

@article{bruninx2014statistical,
  title={A statistical description of the error on wind power forecasts for probabilistic reserve sizing},
  author={Bruninx, Kenneth and Delarue, Erik},
  journal={IEEE Transactions on Sustainable Energy},
  volume={5},
  number={3},
  pages={995--1002},
  year={2014},
  doi={10.1109/TSTE.2014.2320193}
}
}

\end{document}